# Embodied AI Agents for Team Collaboration in Co-located Blue-Collar Work (CAI-BLUE)

Embodied AI Agents for Blue-Collar Work

Kaisa Väänänen

Tauchi Research Centre, Faculty of Information Technology and Communication Sciences, Tampere University, Finland, kaisa.vaananen@tuni.fi

Niels van Berkel

Department of Computer Science, Aalborg University, Denmark, nielsvanberkel@cs.aau.dk

Donald McMillan

Department of Computer and Systems Sciences, Stockholm University, Sweden, donald.mcmillan@dsv.su.se

Thomas Olsson

Tauchi Research Centre, Faculty of Information Technology and Communication Sciences, Tampere University, Finland, thomas.olsson@tuni.fi

Blue-collar work is often highly collaborative, embodied, and situated in shared physical environments, yet most research on collaborative AI has focused on white-collar work. This position paper explores how the embodied nature of AI agents can support team collaboration and communication in co-located blue-collar workplaces. From the context of our newly started CAI-BLUE research project, we present two speculative scenarios from industrial and maintenance contexts that illustrate how embodied AI agents can support shared situational awareness and facilitate inclusive communication across experience levels. We outline open questions related to embodied AI agent design around worker inclusion, agency, transformation of blue-collar collaboration practices over time, and forms of acceptable AI embodiments. We argue that embodiment is not just an aesthetic choice but should become a socio-material design strategy of AI systems in blue-collar workplaces.

# 1 INTRODUCTION

Blue-collar work is fundamentally embodied, and often also highly collaborative and processual. Tasks in manufacturing, maintenance, logistics, and repair are carried out largely through bodily actions, shared material environments, and coordination between co-located workers. Communication is multimodal and situated, relying on tacit knowledge developed through practice [1]. Much of the existing research in collaborative AI has focused on improving the efficiency of white-collar work [2][3], overlooking the physical nature of blue-collar collaboration.

Collaborative AI can take versatile roles at the workplace, including assistants, mediators, and decision supporters (see e.g. [4][5]). The CAI-BLUE[1] project (Collaborative AI for Blue-collar Work, 2026-2029, funded by NordForsk) investigates how AI agents can be designed as collaborative team members rather than as productivity tools for individuals, with an emphasis on responsible, inclusive, and worker-centered design grounded in Nordic workplace values. One of the assumptions in CAI-BLUE is that AI agents should be made visible and physically present in the working environment, to both reduce fear of AI and to advance worker agency, mutual learning, and more broadly, AI literacy [6]. Combining digital interaction with physical artefacts can support spatial and motor skills, reduce cognitive load, and advance natural collaboration [7]. Physical instantiation of user interfaces to technical systems allows for complex social, spatial, and material relationships with available data and affordances [8], which has been shown to support information retrieval and learning [9]. Furthermore, embodied interfaces have been found to support collaborative learning in co-located spaces [10].

To investigate the embodied forms of AI agents in the specific working contexts, we ask: **How can the embodied nature of AI agents support both human–human and human-AI collaboration in co-located blue-collar workplaces?** We envision that embodied AI agents can 1) scaffold shared understanding and coordination, and 2) support inclusive participation and learning across worker experience levels. At the same time, AI embodiments raise open questions around agency, authority, surveillance, and social acceptability that must be addressed through participatory and responsible design. Figure 1 illustrates technology-mediated blue-collar workplace interactions, in a real and in an imagined (AI-generated) situation.

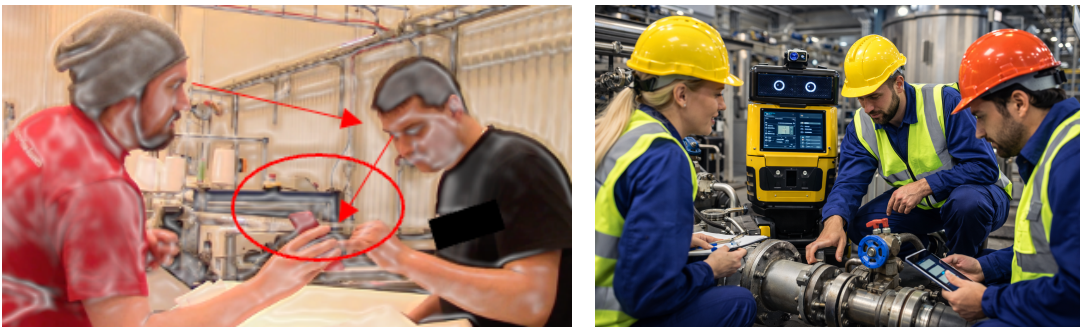

Figure 1: On the left: A real situation in a workplace where two workers are using a mobile device for their communication with the help of AI-based language translation (picture courtesy of Azar Raoufi Masouleh), and on the right, an AI-generated (ChatGPT5.2) situation with an AI agent that has a physically embodied interface.

---

[1] https://www.nordforsk.org/projects/nordic-perspectives-collaborative-ai-blue-collar-work-cai-blue

## 2 SCENARIOS AND BENEFITS OF AI AGENT EMBODIMENTS

The following two speculative scenarios illustrate potential uses of embodied AI agents in a blue-collar workplace:

***Example scenario 1: Learning through collaborative troubleshooting in a smart factory.*** *In a manufacturing facility, a team of assembly line workers collaborates with an AI agent during a shift focused on assembling a new product model. When a mechanical issue arises, the AI assists the team by analysing sensor data and suggesting potential causes. Elina, a mid-career worker, discusses the AI's suggestions with her colleagues, and together they decide on the best course of action. As the team resolves the issue, the agent updates its troubleshooting model based on their chosen solution. This adaptive interaction advances individual learning, while reducing downtime and supporting inclusive decision-making.*
**Benefits of embodying the AI agent:** Rather than addressing individuals through personal interfaces, an embodied agent that is spatially and socially present in the team can explicitly support shared team cognition. By making information visible or audible to all co-located workers simultaneously, the agent supports workers' alignment with the task and each other, without fragmenting their attention or making a single worker as the AI's primary user.

***Example scenario 2: AI transcription for competence development in a car repair shop.*** *Becka and Kalle are working through the annual service for a vehicle, checking its measurements. This collaborative task involves the documentation of a) the measurements themselves, b) supporting documentation, and c) expert recommendations as to when repair or maintenance is needed. Becka has been working as a mechanic for over two decades, while Kalle is newly qualified. The AI transcription system fills out the report by listening to the conversation between the two, dynamically adjusting the final output based on discussion, while also acting as a source of knowledge for Kalle to query workplace practices and regulations at a subsequent time.*
**Benefits embodying the AI agent:** Embodiment may advance the agent's presence in the shared physical space and allows it to participate in the flow of collaborative work by supporting mutual attention. Auditory cues combined with tangible interactions can enable the agent to contribute in ways that feel collaborative rather than interruptive, allowing users to focus on each other and their collaborative task rather than an external system. By building a shared knowledge base based on this natural human-to-human interaction, the AI agent can support situated learning and knowledge sharing while preserving the primacy of human-human communication.

Other benefits of embodied nature of AI agents in collaborative blue-collar work may include:
- Aligning with workers' bodily movements and spatial organisation of work tasks, embodied agents could help smooth transitions and reduce misunderstandings, thereby supporting collaboration indirectly.
- Supporting social cohesion and understanding through the development of shared mental models through physical coordination, whilst maintaining face-to-face interaction patterns.
- Enable hands-free actions and coordination through conversational interfaces, allowing workers to remain focused on physical tasks.

## 3 OPEN QUESTIONS CONCERNING DESIGN OF AI AGENT EMBODIMENT

While embodied AI agents offer promising ways to support human–human collaboration, they also raise critical design and research questions:
**How can embodied agents support worker inclusion without becoming a surveillance instrument?** Many workers are wary of AI systems that monitor performance or worker behaviour. Physical manifestations, such as in embodied agents, may intensify these concerns by subverting the human-centric practices around managing what is visible, audible, and recorded. AI agents should therefore be designed to allow worker control over data collection and provide transparent and clear boundaries around data use.

**How should agency and authority be distributed between workers, management, and AI agents?** Worker autonomy and agency are critical factors in blue-collar work that are at risk when introducing assistive technologies. Embodied agents may be perceived as authoritative or controlling, particularly in hierarchical workplaces. Designers must carefully consider how embodiment, voice, and placement signal the agent's role (e.g. assistant, background resource, or mediator) and how this affects workers' sense of autonomy and responsibility.

**How do embodied AI agents reshape collaboration and work practices over time?** Longitudinal studies are needed to understand how workers adapt to embodied agents, including both their actual work activities and the social connections that are part of their work environment. Special consideration should be given to the potential rise of new and unexpected workplace tensions.

**What forms of embodiment are socially acceptable in blue-collar contexts?** Social acceptability is shaped by cultural norms, safety requirements, and existing power relations. An agent that is acceptable in one setting (e.g. a smart factory) may be intrusive or inappropriate in another (e.g. care or maintenance work). Participatory design with workers is essential to ensure that embodied agents align with local practices and values.

## 4   CONCLUSION

In conclusion, we argue that embodiment is a key factor in designing AI agents that genuinely support human–human collaboration in blue-collar work. Here, embodiment is not an aesthetic choice but a socio-material design strategy. When thoughtfully designed, embodied AI agents can enhance shared understanding, inclusivity, and coordination, while respecting the dignity, expertise, and agency of workers. The challenge for HCI research is to translate these possibilities into responsible, context-sensitive designs that strengthen, rather than disrupt, collaborative work practices. The CAI-BLUE project will address this perspective through the research through design approach, by conducing in-the-wild deployments, and mixed-method evaluations that foreground workers' lived experiences with AI agents.